\documentclass[aps,pra,twocolumn,groupedaddress]{revtex4}
\usepackage{graphicx}

\def\gapx{\lower 2pt \hbox{$\buildrel>\over{\scriptstyle{\sim}}$\ }}
\def\lapx{\lower 2pt \hbox{$\buildrel<\over{\scriptstyle{\sim}}$\ }}
\def\ahop{a_{ho}}
\def\strut{\rule[-0.5cm]{0cm}{1cm}}

\begin{document}
\bibliographystyle{nosrt}

\title{Bose-Einstein condensation in trapped bosons: A Variational Monte Carlo analysis}
\author{J. L DuBois}
\author{H. R. Glyde}
\affiliation{Department of Physics and Astronomy, University of Delaware,
Newark, Delaware 19716,USA}

%\date{Submited PRA: 14 March 2000, Revised \today}

\begin{abstract}
Several properties of trapped hard sphere bosons are evaluated
using variational Monte Carlo techniques.
A trial wave function composed of a renormalized single particle
Gaussian and a hard sphere Jastrow function for pair correlations
is used to study the sensitivity of condensate and 
non-condensate properties to the hard sphere radius and the number 
of particles.  Special attention is given to
diagonalizing the one body density matrix and obtaining the
corresponding single particle natural orbitals and their occupation
numbers for the system.
The condensate wave function and condensate fraction are then
obtained from the single particle orbital with highest occupation.
The effect of interaction on other quantities such as the ground
state energy, the mean radial displacement, and the momentum distribution are 
calculated as well.  Results are compared with Mean Field theory 
in the dilute limit.
\end{abstract}
  %Bose Einstein condensation, 03.75.F

\maketitle

\section{Introduction}

The spectacular demonstration of Bose-Einstein condensation (BEC) in gases of
alkali atoms $^{87}Rb$, $^{23}Na$, $^7Li$ confined in magnetic
traps\cite{anderson95,davis95,bradley95} has led to an explosion of interest in
confined Bose systems. Of interest is the fraction of condensed atoms, the
nature of the condensate, the excitations above the condensate, the atomic
density in the trap as a function of Temperature and the critical temperature of BEC,
$T_c$. The extensive progress made up to early 1999 is reviewed by Dalfovo et
al.\cite{dalfovo99}.

A key feature of the trapped alkali and atomic hydrogen systems is that they are
dilute. The characteristic dimensions of a typical trap for $^{87}Rb$ is
$a_{h0}=\left( {\hbar}/{m\omega_\perp}\right)^\frac{1}{2}=1-2 \times 10^4$
\AA\ (Ref. 1). The interaction between $^{87}Rb$ atoms can be well represented
by its s-wave scattering length, $a_{Rb}$. This scattering length lies in the
range $85 < a_{Rb} < 140 a_0$ where $a_0 = 0.5292$ \AA\ is the Bohr radius
\cite{gardner95}. The definite value $a_{Rb} = 100 a_0$ is usually selected and
for calculations the definite ratio of atom size to trap size 
$a_{Rb}/a_{h0} = 4.33 \times 10^{-3}$ 
is usually chosen \cite{dalfovo99}. A typical $^{87}Rb$ atom
density in the trap is $n \simeq 10^{12}- 10^{14}$ atoms/cm$^3$ giving an
inter-atom spacing $\ell \simeq 10^4$ \AA. Thus the effective atom size is small
compared to both the trap size and the inter-atom spacing, the condition
for diluteness (i.e., $na^3_{Rb} \simeq 10^{-6}$ where $n = N/V$ is the number
density). In this limit,
although the interaction is important, dilute gas approximations such as the
Bogoliubov theory\cite{bogoliubov47}, valid for small $na^3$ and large
condensate fraction $n_0 = N_0/N$, describe the system well. Also, since most
of the atoms are in the condensate (except near $T_c$), the Gross-Pitaevskii
equation\cite{gross61,pitaevskii61} for the condensate describes the whole gas
well. Effects of atoms excited above the condensate have been incorporated
within the Popov approximation\cite{hutchinson97}. One of the chief purposes of
this paper is to go beyond the dilute limit, to test the limits of these
approximations and to explore the properties of the trapped Bose gas as $na^3$
increases between the dilute limit and the dense limit.  We  use Variational Monte
Carlo (VMC) methods. We increase the density by increasing both $N$ and the s-
wave scattering length up to the value $na^3 \simeq 0.21$ which 
describes liquid $^4$He at SVP when the $^4$He atoms are represented by hard
spheres of diameter $a = 2.203$ \AA\cite{kalos74}.

In addition to the mean-field theories noted above, the trapped Bose gas at
finite temperatures has been investigated using Path Integral Monte Carlo (PIMC)
methods. Krauth\cite{krauth96} simulated 10,000 atoms in a spherical trap with
ratio of scattering length $a$ to trap length given by 
$a/a_{h0}=4.3\times10^{-3}$ noted
above. He showed that the critical temperature $T_c$ is lowered compared to the
ideal Bose gas as a result of interaction. $T_c$ is lower because the repulsion
between the atoms spreads the atoms in the trap and lowers the density compared
to the non- interacting case. The same result has been obtained in mean-field
approximations \cite{dalfovo99}. Krauth also showed that, in the dilute limit,
the condensed atoms are highly concentrated at the center of the trap while the
uncondensed or thermal atoms are spread out over a wide range, are dilute and
well approximated by a classical, ideal gas. There is little interaction between
the condensed and uncondensed components (see also \cite{dalfovo99}).

Gr\"{u}ter et al.\cite{gruter97} evaluated $T_c$ for a uniform, bulk, hard
sphere Bose gas over a wide density range, from dilute to liquid $^4$He
densities. They find that $T_c$ is increased above the ideal Bose gas value by
interaction in the dilute range. In the uniform gas case the density is not changed
by interactions. At liquid $^4$He densities, $T_c$ is decreased by
interaction\cite{pollock92,wilks67}.

Holzmann et al.\cite{holzmann99} made a direct comparison between Hartree-Fock
(HF) and PIMC calculations of the number density $N(r)$ of atoms in a
trap. The atoms were again represented by hard spheres with $a/a_{h0}=0.0043$.
From $N(r)$, they find for temperatures near $T_c$ that the condensate
fraction $N_0$ is larger in the exact PIMC evaluation than in the HF
approximation. The energy beyond the Hartree-Fock approximation is often denoted
the ``correlation'' energy. This correlation apparently allows $N_0$ to
increase at a given $T$ and allows condensation to begin at a higher
temperature. At lower temperature $T\ \lapx 0.75 T_c$, there is excellent
agreement between the PIMC and HF $N(r)$. The increase in $N_0$ with exact
representation of the interaction effects is consistent with the corresponding
increase in $T_c$ with interaction in the uniform Bose gas.

Giorgini et al.\cite{giorgini} have evaluated the ground state energy $E/N$ and
the condensate fraction $N_0/N$ at $T=0$ K of a uniform Bose gas over a wide
density range $(10^{-6} \leq na^3 \leq 10^{-1})$ using Green Function Monte Carlo
(GFMC) methods. They find that the mean field results of $E/N$ and the
Bogoliubov result for $N_0/N$ agree well with the GFMC values in the density
range $10^{-6} \leq na^3 \leq 10^{-3}$. However, there are clear differences at 
higher densities $na^3 \geq 10^{-3}$ (helium density is $na^3 \simeq 0.21$). The
results are not sensitive to reasonable variation of the inter-boson
potential\cite{giorgini}.

In this context, we have evaluated the ground state properties of
a trapped, hard sphere Bose gas over a wide range of densities
using Variational Monte Carlo (VMC) methods. We begin in the
dilute limit, small $N$ and $a/a_{ho}=0.0043$ corresponding to
$^{87}Rb$ in a trap, and increase both $N$ and $a$ separately to
increase the density up to liquid $^4$He densities. At the lower
densities we compare the energy $E/N$ and root mean square
amplitudes $<x^2+y^2>$ and $<z^2>$ of atoms in an anisotropic trap
with Gross-Pitaevskii (GP) results\cite{dalfovo96}. The two
methods agree well at low densities but even at densities $na^3
\approx 10^{-5}$ small differences in $E/N$ are readily apparent. 
Also, at higher densities we find the effects of interaction depend
separately on $N$ and $a/a_{ho}$, not simply in the product 
$Na/a_{ho}$ as appears in the GP theory. As density is increased 
still further,
we find that the condensate is no longer concentrated at the
center of the trap. Rather increased interaction (increased $a/a_{ho}$)
depletes the condensate at the center and the condensate appears
at the edges of the trap -- as found in liquid $^4$He droplets
(Lewart et al.\cite{lewart88}). Also, as density increases,
correlations in the single particle density appear, reflecting the
interaction in a confined space, in the same way that interaction
at higher density introduces correlation in the pair correlation
function in the uniform case. We evaluate the momentum
distribution and condensate fraction over the dilute to dense
range and compare with mean-field results\cite{javanainen96}.

In section 2 we introduce the Hamiltonian, the wave function and the MC method
and the definition of the natural orbitals.  The results are presented in 
section 3 and discussed in section 4.

\section{Bosons in a Harmonic Trap}
\subsection{The System}
We consider N bosons of mass $m$ confined in an external 
trapping potential, $V_{ext}({\bf r})$, and interacting via a two-body 
potential $V_{int}({\bf r}_1,{\bf r}_2)$.  
The Hamiltonian for this system is:
\begin{equation}
    H = \sum_i^N \left(
        \frac{-\hbar^2}{2m}
        { \bigtriangledown }_{i}^2 +
        V_{ext}({\bf{r}}_i)\right)  +
        \sum_{i<j}^{N} V_{int}({\bf{r}}_i,{\bf{r}}_j).
\end{equation}
We consider both a 
spherically symmetric (S) harmonic trap and an elliptical (E) harmonic trap,
\begin{equation}
V_{ext}({\bf r}) = 
\Bigg\{
\begin{array}{ll}
        \frac{1}{2}m\omega_{ho}^2r^2 & (S)\\
\strut
	\frac{1}{2}m[\omega_{ho}^2(x^2+y^2) + \omega_z^2z^2] & (E)
\label{trap_eqn}
\end{array}
\end{equation}
Here $\omega_{ho}^2$ defines the trap potential strength.  In the case of the
elliptical trap, $V_{ext}(x,y,z)$, $\omega_{ho}=\omega_{\perp}$ is the trap frequency
in the perpendicular or $xy$ plane and $\omega_z$ the frequency in the $z$
direction.
The mean square vibrational amplitude of a single boson at $T=0K$ in the 
trap (\ref{trap_eqn}) is $<x^2>=(\hbar/2m\omega_{ho})$ so that 
$a_{ho} \equiv (\hbar/m\omega_{ho})^{\frac{1}{2}}$ defines the 
characteristic length
of the trap.  The ratio of the frequencies is denoted 
$\lambda=\omega_z/\omega_{\perp}$ leading to a ratio of the
trap lengths
$(a_{\perp}/a_z)=(\omega_z/\omega_{\perp})^{\frac{1}{2}} = \sqrt{\lambda}$.

We represent the inter boson interaction by a pairwise, hard core potential
\begin{equation}
V_{int}(r) =  \Bigg\{
\begin{array}{ll}
        \infty & {r} \leq {a}\\
        0 & {r} > {a}
\end{array}
\end{equation}
where ${a}$ is the hard core diameter of the bosons.  Clearly, $V_{int}(r)$
is zero if the bosons are separated by a distance $r$ greater than $a$ but
infinite if they attempt to come within a distance $r \leq a$.

The weak interaction limit is $a << a_{ho}$
and $a << n^{-\frac{1}{3}}$ (where $n=N/V$ is the local number density), 
a hard core diameter small compared to
the dimensions of the trap and compared to the inter-particle spacing
$l=(V/N)^{\frac{1}{3}}$.  
For trapped alkali atoms we have typically $na^3 \lapx 10^{-5}$.  
Introducing lengths in units of $a_{ho}$, $r \rightarrow r/a_{ho}$, 
and $\hbar\omega_{ho}$ as units of energy as in \cite{dalfovo99},  
the Hamiltonian is:
\begin{equation}
H = \sum_i^N\frac{1}{2}(-\bigtriangledown _{i}^2 + x_i^2+y_i^2 + \lambda^2 z_i^2) +
    \sum_{i<j} V_{int}(|{\bf r}_i-{\bf r}_j|).
\end{equation}
Since there 
is Bose condensation, we have $n\lambda^{3}_T \gapx 2.616$ where $\lambda_T$
is the atomic thermal wavelength.  Thus we are in the regime where the atomic
wavelength is long compared to the hard core diameter, $\lambda_T >> a$ or
$ka << 1$ where $k \equiv 2\pi/\lambda_T$.
The scattering of two particles interacting via a hard core potential in the
limit $ka << 1$ is purely $s$ wave with scattering length $a$.  If we
approximate the full potential between the two particles by a contact 
potential,
\begin{equation}
v(r) = g\delta({\bf r}) = \frac{4\pi\hbar^2 a}{m}\delta({\bf r})
\end{equation}
the scattering length in this limit between the two is again purely $s$ wave
with scattering length $a$.  Thus we may compare directly results calculated
using a hard core potential (3) and with a contact potential approximation (5)
in the regime $a << \lambda_T$.  Specifically, we may compare the present MC
results with results calculated using (5) and the mean field, Gross-
Pitaevskii (GP) equation.  This comparison is especially interesting in the
dilute limit $na^3 <<1$.  At high densities $(na^3 \gapx 0.1)$, we expect
short range, pair correlations induced by the hard core to be important and
short range correlations are not well described by a mean field theory.

\subsection{Wave Function}
To describe the ground state of the $N$ bosons, we introduce a variational
trial wave function which is a product of a single particle
function $g({\bf r})$ and a pair Jastrow function \cite{jastrow} 
$f(|{\bf r}_1-{\bf r}_2|)$,
\begin{equation}
\Psi_\nu({\bf r}_{_1}..{\bf r}_{_N},\alpha,\beta) =
  \prod_{i}g(\alpha,\beta,{\bf r}_i)
  \prod_{i<j} f(a,|{\bf r}_i-{\bf r}_j|)
  \label{wvfn}
\end{equation}
where $\alpha$ and $\beta$ are the variational parameters.  
We select a single particle function, 
\begin{equation}
g(\alpha,\beta,{\bf r}_i) =
    \exp \left[-\alpha(x^2_i+y^2_i+\beta z^2_i)\right]
\end{equation}
which is a HO ground state function having two variational parameters,
$\alpha$ and $\beta$.  For spherical traps, $\beta=1$, and for noninteracting
bosons $(a=0)$, $\alpha=1/2a_{ho}^{2}$.  For the pair function we select
the exact solution
of the Schroedinger equation for two particles interacting via the hard core
potential (3) in the limit $k\rightarrow0$, i.e.
\begin{equation}
f(a,r) = \left\{
\begin{array}{ll}
    (1-\frac{a}{r}) & r > a \\
    0 & r \leq a
\end{array}
\right.
\label{jastrow_func}
\end{equation}
(see for example Huang \cite{huang63}). The $\Psi_\nu({\bf r}_{_1}..{\bf r}_{_N})$ therefore has the correct
form for small $|{\bf r}_i-{\bf r}_j|$ and has two variational
parameters $\alpha$ and $\beta$ that describe the spread of the 
bosons in the trap as the hard core diameter is increased.
By constructing the wave function in this way, we limit the number of 
variational parameters while preserving the correct functional form 
in the $a \rightarrow 0$ limit.  However, the lack of any variational parameters
in the Jastrow term is a potential source of inaccuracy. 

We then minimize the expectation value of the Hamiltonian as obtained from
$$ <H> = \frac {\int{d{\bf r}_1..d{\bf r}_N\Psi^*_{\nu}\Psi_{\nu}
(\frac{ {\it H}\Psi_{\nu} }{ \Psi_{\nu} }) } }
   {\int{d{\bf r}_1..d{\bf r}_N\Psi^*_{\nu}\Psi_{\nu}}}
$$
with respect to $\alpha$ and $\beta$ using the Metropolis Monte Carlo 
method of integration.
This is accomplished by using a Metropolis random walk \cite{metropolis}
to generate a set of $M$ particle configurations,
$\Omega_1..\Omega_M$ which
conform to the probability distribution $|\Psi_\nu|^2$. We then approximate
$<H>$ by summing over the `local energy' as follows
$$
<H> \gapx \frac{1}{M}\sum^M_{n=1} \left(\frac{H(\Omega_n)\Psi_\nu(\Omega_n)}
{\Psi_\nu(\Omega_n)}\right)
$$
For a review of the Variational Monte Carlo method, see \cite{vmc_Review}.

\subsection{Condensate and Natural Orbitals}
A goal here is to calculate the condensate fraction and condensate density
in the ground state.  To do this we require a definition of the condensate
single particle state.  Following Penrose and Onsager, L\"owdin and others
\cite{onsager56,lowdin55}, we take the one-body density matrix (OBDM)
as the fundamental quantity for an interacting system and define the 
natural single particle orbitals (NO) in terms of the OBDM.

The OBDM is \cite{baym76}
\begin{equation}
\rho({\bf r'},{\bf r}) = <\hat{\Psi}^{\dagger}({\bf r}'),\hat{\Psi}({\bf r})>,
\end{equation}
where $\hat{\Psi}({\bf r})$ is the field operator that annihilates 
a single particle
at the point ${\bf r}$ in the system.  At $T=0K$, the expectation value is
evaluated using the wave function  $\Psi_\nu({\bf r})$ in (6).
To define the NO, we introduce a set of single particle states having wave
functions $\phi_i({\bf r})$ and expand $\hat{\Psi}({\bf r})$ in terms of the 
operators $\hat{a}_i$ which annihilate a particle from state $|i>$,
\begin{equation}
\hat{\Psi}_0 =\sum_i\phi_i({\bf r})\hat{a}_i.
\end{equation}
Requiring that the $\hat{a}_i$ satisfy the usual commutation 
$([\hat{a}_i^{\dagger},\hat{a}_j] = \delta_{ij})$ and number relations
$(<\hat{a}_i^{\dagger}\hat{a}_j>=N_i\delta_{ij})$, we have

\begin{equation}
\begin{array}{r}
\rho({\bf r},{\bf r}') =
        \sum_{ij}\phi_j^{*}({\bf r}')\phi_i({\bf r})N_i\delta_{ij}\\
        =\sum_{ij}\phi_j^{*}({\bf r})\phi_i({\bf r}')N_i\delta_{ij}
\end{array}
\end{equation}
This may be taken as the defining relation of the NO, $\phi_i({\bf r})$.
Specifically, we have from (11),

\begin{equation}
        \int{
                d{\bf r}..d{\bf r}'
		\phi_i^{*}({\bf r})
		\rho({\bf r},{\bf r}')
		\phi_j({\bf r}')
		= N_i\delta_{ij}
        },
\end{equation}
so that the NO may be obtained by diagonalizing the OBDM.  The eigenvectors
are the NO and the eigenvalues are the occupation, $N_i$, of the orbitals.  The 
condensate is the orbital having the highest occupation, denoted $\phi_0(r)$
and the condensate fraction is $n_0=N_0/N$.

The relations (11) and (12) involve the vector ${\bf r}$ and ${\bf r}'$
and cannot be solved directly as matrix equations.  To obtain matrix
equations, we restrict ourselves to spherical traps and seek equations for the
radial component of the NO. Assuming the potential seen by a single particle,
including inter-particle interaction, is spherically symmetric, the NO will
have the form 
\begin{equation}
	\phi_i({\bf r}) = R_{nl}(r)Y_{lm}(\theta,\phi)
\end{equation}
Where $Y_{lm}(\Omega)$ are the spherical harmonics and $R_{nl}(r)$ is the
radial wave function and $i=n,l,m$ are the state indices.
We expand the OBDM in its angular momentum $(l)$ components $\rho_l(r,r')$
as
\begin{equation}
\rho({\bf r},{\bf r}') = \sum_l \frac{(2l+1)}{4\pi}P_l(\hat{\bf r}
\cdot\hat{\bf r}')
\rho_l(r,r')
\end{equation}
where $P_l(cos\theta)$ are Legendre Polynomials in the angle between
$\hat{\bf r}$ and $\hat{\bf r}'$ and
\begin{equation}
\begin{array}{l}
\rho_{_l}(r_{_1},r_{_1}') =\\
\\
        \int{
                d\Omega_1d{\bf r}_{_2}..d{\bf r}_{_N}
                \Psi^{*}({\bf r}_{_1}..{\bf r}_{_N})
                P_l(\hat{r}_{_1}\cdot\hat{r}'_{_1})
                \Psi({\bf r}'_{_1}..{\bf r}_{_N})
        }.
\end{array}
\end{equation}
Substituting (13) and (14) into (11) and using the properties of the
spherical harmonics,
$Y_{lm}(\Omega)$, we obtain a relation equivalent to (11) for each
$l$ component $\phi_{nl}(r)$ of the NO,
\begin{equation}
\rho_{_l}(r,r') = \sum_n \phi_{nl}(r)\phi_{nl}(r')N_{nl}
\end{equation}
where $i=nl$.  To solve this equation readily as a matrix equation, we
introduce the radial function
\begin{equation}
u_{nl}(r) = r\phi_{nl}(r).
\end{equation}
The $u_{nl}(r)$ is well behaved at $r \rightarrow 0$ and has dimensions
$L^{-\frac{1}{2}}$ like a one-dimensional wave function.  
In terms of the $u_{nl}(r)$,
the defining relations are:
\begin{equation}
[r\rho_{_l}(r_{_1},r_{_1}')r'] = \sum_n u_{nl}(r)u_{nl}^{*}(r')N_{nl}
\end{equation}
and
\begin{equation}
\int dr [r\rho_{_l}(r_{_1},r_{_1}')r'] u_{nl}(r') = u_{nl}(r)N_{nl}
\end{equation}
The $[r\rho_{_l}(r_{_1},r_{_1}')r']$ serves as a one-dimensional OBDM along the radial
coordinate.  This $1D$ matrix relation may be solved numerically on a grid
in $r$ to obtain the $u_{nl}(r)$ as eigenvectors and the $N_{nl}$ as
eigenvalues.  The condensate orbital is 
\begin{equation}
	\phi_0(r) = \frac{1}{\sqrt{4\pi}}u_{00}(r)/r.
\end{equation}

The momentum distribution may be obtained from the OBDM as well given that
the orbitals in momentum space are 
\begin{equation}
\tilde{\phi}_i({\bf k}) = \frac{1}{ (2\pi)^{\frac{3}{2}} }
			\int d{\bf r} e^{ -i{\bf k}\cdot{\bf r} } 
			{\phi}_i({\bf r})
\label{phi_k}
\end{equation}
and the momentum distribution is
\begin{equation}
\tilde{\rho}({\bf k}) = \sum_i n_i|\tilde{\phi}_i({\bf k})|^2.
\label{rho_k}
\end{equation}
Substituting (\ref{phi_k}) into (\ref{rho_k}) obtains an expression for
$\tilde{\rho}({\bf k})$ in terms of the OBDM:
\begin{equation}
\tilde{\rho}({\bf k}) = \frac{1}{(2\pi)^3}\int d{\bf r}d{\bf r}'
		\rho({\bf r},{\bf r}')e^{i{\bf k}\cdot 
		({\bf r}-{\bf r'})}.
\end{equation}
The momentum distribution for a spherically symmetric system is therefore 
\begin{equation}
\begin{array}{l}
{\displaystyle
\tilde{\rho}({\bf k}) = \tilde{\rho}(|{\bf k}|) = 
}\\ 
\strut
\strut
{\displaystyle
\frac{1}{2\pi^2}\int \!\!d{\bf r_1}..d{\bf r_N}\  
	|\Psi_{\nu}({\bf r}_1..{\bf r}_N)|^2
}\\
{\displaystyle
\times	\int_0^{\infty}\!\!\!\!dr \  r^2 \frac{sin(kr)}{kr}
	\frac{\Psi_{\nu}({\bf r}_1\!\!+\!{\bf r},{\bf r}_2..{\bf r}_N)}
	{\Psi_{\nu}({\bf r}_1..{\bf r}_N)}
}
\end{array}
\end{equation}
where the orientation of ${\bf r}$ may be chosen arbitrarily.
\section{Results}

    \begin{figure}
\begin{center}
\includegraphics[width=3in]{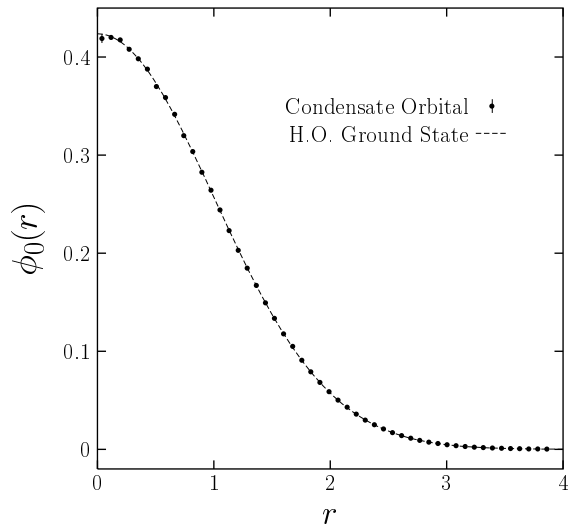}
\caption{ \footnotesize
    The condensate orbital, $\phi_0(r)$, obtained by numerical
    diagonalization of the OBDM $\rho_0(r,r') = \sum_i n_i \phi_i^*(r)\phi_i(r')$
    calculated by Variational Monte Carlo (VMC) (solid dots) for an ideal
    Bose gas in a harmonic trap.  The dashed line is the harmonic oscillator (HO)
    ground state wave function for the same trap.  $\phi_0(r)$ and $r$ are 
    dimensionless in units of $a_{ho}$.}
\end{center}
    \end{figure}

    \begin{figure}[t]
\begin{center}
\includegraphics[width=3in]{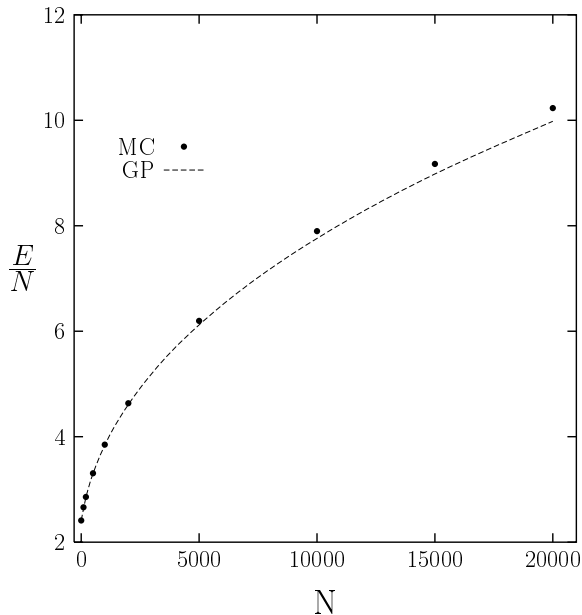}
\caption{\footnotesize Energy per particle, 
	in units of $\hbar\omega_{\perp}$, for
        hard sphere bosons in an anisotropic trap
        as a function of the number of particles N in the trap.  
	Solid circles are the
        present VMC values for hard spheres with diameter corresponding to
        the scattering length of Rb, $a_{Rb} = 4.33 \times 10^{-3} \ahop$ --
        where $\ahop$ is the trap length in the perpendicular direction.
        The error bars lie within the solid dots.  The dashed line is 
	the value obtained using the Gross-Pitaevskii equation
        (GP) for the same system \cite{dalfovo96}.}
\end{center}
    \end{figure}

    \begin{figure}[th!]
\begin{center}
\includegraphics[width=3.2in]{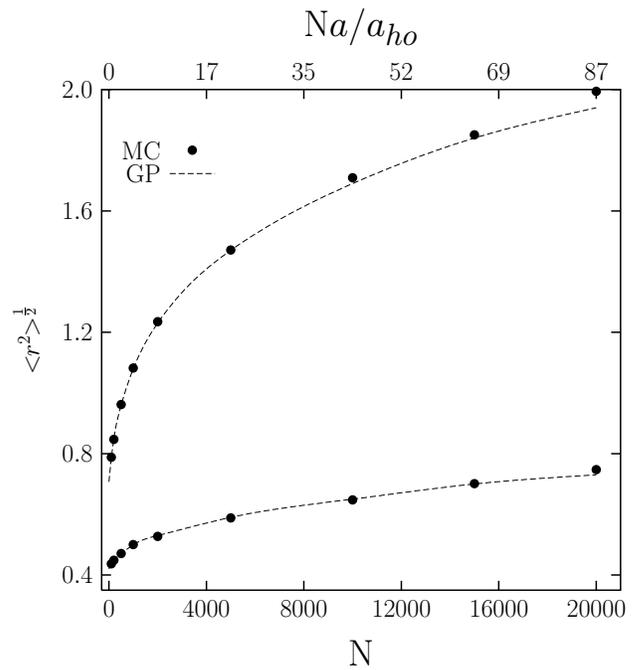}
 \caption{\footnotesize
    Average axial displacement of bosons from the
    center of an anisotropic trap in the $\perp$ direction, $<\!\!x^2\!\!+\!\!y^2\!\!>^{\frac{1}{2}}$,
    (top) and in the $z$ direction $<\!\!z^2\!\!>^{\frac{1}{2}}$  (bottom) expressed in
    units of the perpendicular trap length $\ahop$
    as a function of the number of particles, $N$.
    Solid dots are from the present VMC results 
    for hard spheres with diameter
    $a_{Rb} = 4.33 \times 10^{-3} \ahop$ .
    The dashed lines are the values obtained from the Gross-Pitaevskii
    equation (GP) for the same $a_{Rb}$ \cite{dalfovo96}. }
\end{center}
    \end{figure}
In this section we present the results of the present
Variational Monte Carlo (VMC) calculation of properties of bosons at $T = 0$ K
confined in a harmonic trap. To begin, Fig. 1 shows the condensate orbital
$\phi_0(r)$ for independent, non-interacting bosons in a spherical harmonic trap
compared to the harmonic oscillator (HO) ground state function for the same
trap. The $\phi_0(r)$ is calculated by evaluating the one body density matrix
(OBDM) and diagonalizing it numerically to obtain the single particle orbitals
and their occupation as discussed in section 2. For no interaction, only
$\phi_0(r)$ is occupied $(n_0 = 1)$. The excellent agreement between $\phi_0(r)$
and the HO ground state function for all $r$ provides a good check of the method
for non-interacting bosons. The statistical sample is proportional to $r^2$ and
the statistics become poor at $r \rightarrow 0$.

Fig. 2 shows the energy per particle, $E/N$, of a gas of $N$ weakly interacting
hard sphere bosons in an ellipsoidal trap. 
The ratio of the characteristic length of the short axis ($z$ axis) to the longer
perpendicular ($x$ and $y$) axis of the trap is $(a_z/z_\perp) = 1/\sqrt{\lambda}$;, 
$\lambda =\sqrt{8}$. 
The hard sphere diameter, $a$, (scattering length) of the bosons
corresponds to $^{87}Rb$ atoms in an ellipsoidal trap with
$a/a_{ho} = 0.00433$ with $a_\perp = a_{ho}$. 
In Fig. 2 the dots
are the present VMC $E/N$ of the whole Bose gas while the dashed line is the
$E/N$ of the condensate calculated by Dalfovo and Stringari\cite{dalfovo96}
using the Gross-Pitaevskii (GP) equation.  Our purpose is to make comparisons
between the present VMC calculation and GP equation results across the
dilute regime for which GP is expected to be valid.  The region $100 \leq N
\leq 20000$ corresponds to atom densities at the center of the trap of $2 \times
10^{-6}\ \lapx na^3\ \lapx 2 \times 10^{-5}$. In the earliest
experiments\cite{anderson95} with $^{87}Rb$ there were typically $N = 10,000$
atoms in the trap. In more recent experiments $N$ is larger, $N \simeq 10^5 -
10^6$.

At small $N$ values in the dilute limit, 
there is excellent
agreement between the GP and VMC energies. In this regime, the mean-field, GP
equation is expected to be accurate. 
As N increases a clear difference between the VMC and GP $E/N$
values emerges. We find that for a fixed scattering length, 
the difference $\delta(E/N) = (E_{MC}-E_{GP})/N$ is proportional to $N^{\frac{3}{5} }$ and 
can be well represented as
\begin{equation}
\delta(E/N) = [(3.0\times10^{-5} \pm 10^{-6}) \times (N)^{\frac{3}{5}} - 0.052]\hbar\omega_{\perp}.
\end{equation}
The difference is $1.8\%$ at $N=10,000$ and $2.5\%$ at $N=20,000$.
The difference, we believe, arises largely because there is excitation of bosons above
the condensate. $E_{MC}/N$ includes the excited atoms while
$E_{GP}/N$ is the energy of the condensate alone.
We return to this point in the Discussion.

Fig.~3 compares the root mean square displacement of hard sphere bosons from the
center of the same anisotropic trap discussed in Fig.~2 calculated using VMC and
the GP equation. The upper (lower) line is the radial displacement along the
$\perp (z)$ direction. The agreement between the GP and the VMC displacements is
excellent, right up to $N = 20,000$ for $a = a_{Rb}$.

It is interesting that the VMC and GP displacements agree well while the
energies in Fig.~2 differ for $N \gapx 10^4$. Essentially, the $E/N$ is very
sensitive to the few high energy bosons at the edges of the trap. That is, the
small number of atoms having large displacements increase the energy
significantly but change $<r^2>$ little. An example of this effect is found in
the Thomas-Fermi approximation where the density is cut off to zero at a
specific radius \cite{dalfovo99}. In the TF model there is no tail in the density reaching up to
large $r$ values. The cut off changes $<r^2>$ little but $E/N$ is
significantly affected \cite{dalfovo99}.

Fig.~4 again shows $E/N$ for hard sphere bosons in an anisotropic trap as a
function of $Na/a_{ho}$. 
However, in this case the product $Na/a_{ho}$ is adjusted by varying
both $N$ and $a/a_{ho}$. 
The star symbols show $E/N$ for $a=a_{Rb}$ and $2000 \leq N \leq 20,000$, the
crosses for $a=10a_{Rb}$ and $200 \leq N \leq 2,000$ and the square
$a=20a_{Rb}$ and $100 \leq N \leq 1,000$.
If the impact of
interaction and $E/N$ depended solely on the product $Na/a_{ho}$, as is the case in the
mean-field, GP equation, all three lines in Fig.~4 would coincide. $E/N$ clearly
depends separately on $N$ and $a/a_{ho}$, even in the region of $N = 10,000-20,000$.
The separate dependence is not large at $Na/a_{ho} \approx 20$ but becomes increasingly
large as $Na/a_{ho}$ increases. Also, at these and larger densities, the parameter
which determines the magnitude of corrections arising from interactions is
$N^\frac{3}{5}(a/a_{ho})^{\frac{8}{5} }$. Apparently the interaction effects depending on $Na/a_{ho}$ are
valid only in the limit of small $a$ ($a/a_{ho} << 1$).  We examine the functional form of
the separate dependence of $E/N$ on $N$ and $a/a_{ho}$ in the discussion section.
    \begin{figure}
\begin{center}
\includegraphics[width=3in]{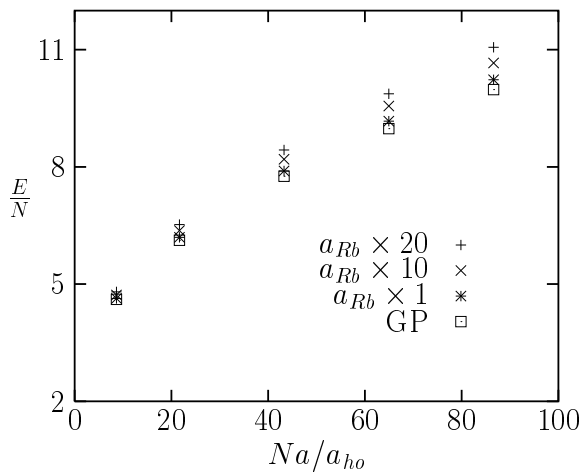}
\caption{\footnotesize 
	Energy per particle, in units of $\hbar\omega_{\perp}$, for
        interacting bosons in an anisotropic trap
        as a function of the product N$a/a_{ho}$ (number of particles N and scattering
	length $a$) as obtained by the present VMC calculation for hard spheres
	with core diameter $a = 1,10,20$ times the scattering length of $^{87}$Rb,
	$a_{Rb} = 4.33 \times 10^{-3}a_{ho}$.  Open squares are Gross-Pitaevskii 
	equation results for the same N$a$.  Estimated errors lie within symbol size. 
        }
\end{center}
    \end{figure}

Having investigated lower densities and made comparisons with results obtained
using the GP equation, we now turn to higher densities and bosons represented by
hard spheres having larger hard core diameters. We evaluate the OBDM and the
density for these cases going up to densities comparable to liquid $^4$He
droplets. Fig. 5 shows the condensate orbital (wave function) for 128 bosons in
a spherical harmonic trap as the hard core radius, $a$, is increased from zero
up to $a = 64 a_{Rb}\ (a_{Rb}/a_{ho} = 0.00433)$. The case $a/a_{Rb} = 1$
corresponds to $N = 128$ Rb atoms in a spherical trap. Clearly the condensate
orbital spreads out in the trap as $a$ increases. At $a/a_{Rb} = 64$, the
condensate density is effectively constant in the trap out to nearly three times
the trap length parameter $a_{ho}$. For these larger core radii, the appropriate
measure of the interaction is $na^3$ where $n = N/V$. For $a = 64 a_{Rb}$, $na^3
\simeq 2 \times 10^{-3}$ at the center of the trap.
    \begin{figure}
\begin{center}
\includegraphics[width=3in]{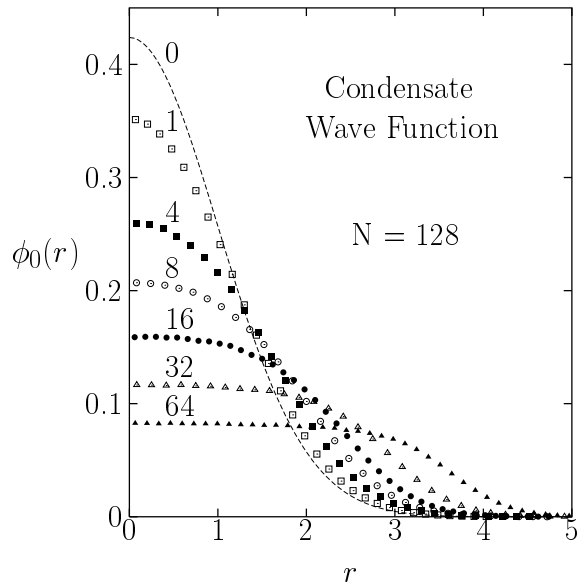}
\caption{\footnotesize
Condensate natural orbital $\phi_0(r)$ for 128 hard spheres with core diameter 
$a = 0,1,...,64$ times the scattering length of
$^{87}$Rb, $a_{Rb} = 4.33 \times 10^{-3}a_{ho}$. 
in a spherically symmetric harmonic trap 
All lengths are in 
terms of the trap length $a_{ho} = \sqrt{{\hbar}/{m \omega_{ho}}}$.
        }
\end{center}
    \end{figure}

In Fig. 6 we show the total density $\rho(r)$ and the density of atoms in the
condensate orbital as $N|\phi_0(r)|^2$ for 64 bosons in a spherical trap with
$a$ increased to $128 a_{Rb}$, $256 a_{Rb}$ and $512 a_{Rb}$. Since we plot
$N|\phi_0(r)|^2$ rather than $N_0|\phi_0(r)|^2$, ``condensate density'' can
exceed the total density. At $a=256a_{Rb}$ the condensate is $n_0=N_0/N = 20\%$.
We note that as $a$ increases, the condensate moves
away from the center of the trap. At $a = 512 a_{Rb}$, the condensate is at the
edges of the trap.  When $na^3$ is large at the center of the trap, the maximum
condensate density is in the region
of lower particle density found at the edge of the trap, as calculated for 
liquid $^4$He droplets
\cite{lewart88}. Thus the location of the condensate is entirely different at
small and large scattering length. In a slave boson approach, depletion of the
condensate at the center of the trap has also been demonstrated 
\cite{ziegler97}.
    \begin{figure}[t!]
\center
\includegraphics[height = 4in]{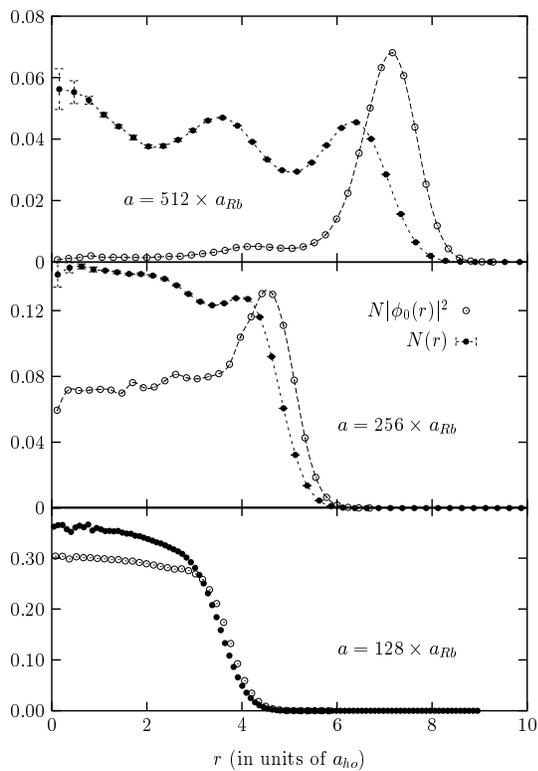}
 \caption{\footnotesize
    Condensate density, $|\phi_0(r)|^2$, scaled by the number of particles $N$,
    vs. total particle density, $N(r)$, with scattering length
    $a = 128,256,512 \times a_{Rb}, $ from bottom to top -- where $N(r)$ is
    normalized to $N$, $|\phi_0(r)|^2$ is normalized to $1$, and all lengths
    are in units of the characteristic trap length $a_{ho}$.
        }
    \end{figure}
Also, at large $a/a_{ho}$ the total density develops correlations. These correlations
reflect the inter-boson correlations induced by the hard core interaction. For
$a = 512 a_{Rb}$ we have $a/a_{ho} \simeq 2.2$. With a peak
in the density at $r = 0$, we expect the first minimum in the density at $r
\simeq a/a_{ho} \simeq 2.2$ as in the upper frame of Fig. 6.

Fig. 7 gives the fraction of bosons in the condensate orbital, $n_0$, calculated
by VMC and diagonalization of the OBDM corresponding to the condensate orbitals
shown in Fig. 5. The $n_0$ values, as in Fig. 5, are for 128 hard sphere bosons
in a spherical trap with hard sphere radius $a/a_{Rb} = 1, 2, 4, 16, 32$ and 64
e.g. $a = 64 \times a_{Rb} = 64 \times 4.33\times10^{-3}\ a_{ho} = 0.277 a_{ho}$. The
Bogoliubov \cite{bogoliubov47} result for $n_0$ adapted to a spherical
trap\cite{javanainen96} is shown as a dashed line. Visible departures of the
VMC $n_0$ from the Bogoliubov result begin at $n_0 \simeq 0.96$ (a depletion of
4\%) corresponding to a density at the center of the trap $na^3 \simeq 1 
\times 10^{-3}$. In the uniform gas case, the Bogoliubov result remains accurate
up to a condensate fraction $n_0 \simeq 0.89$ (11\% depletion) which occurs at a
density $na^3 \simeq 5 \times 10^{- 3}$ \cite{giorgini}. The Bogoliubov
approximation has a more limited range of application for bosons in a trap
because the interaction changes the density profile (and the shape of the
condensate wave function) as well as simply depleting the condensate and $n_0$
depends on the density and shape of the condensate wave function. In the uniform
case, the density cannot change.
    \begin{figure}
\begin{center}
\includegraphics[width = 3.1in]{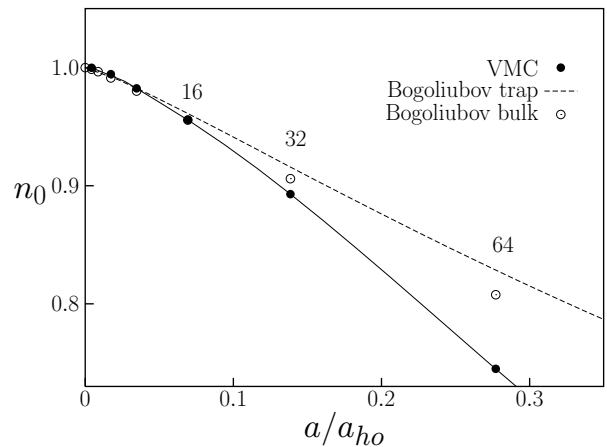}
 \caption{\footnotesize
The condensate fraction, $n_0$, for 128 particles in
a harmonic trap as a function of the ratio of scattering length to
trap length $a/a_{ho}$ using three
methods.  Occupation numbers for the ground state orbital $\phi_0$
found using Variational Monte-Carlo (MC) methods (solid dots) agree well with
values of $n_0$ obtained from the Bogoliubov equations for a uniform gas (open circles) 
and a Mean Field Bogoliubov approximation (dashed line) \cite{javanainen96}
for $n_0 > .9$.
        }
\end{center}
    \end{figure}
In Fig 8., we again show the condensate fraction, $n_0$, as a function of the
ratio of the scattering length, $a$, to the characteristic trap length,
$a_{ho}$.  Here, $n_0$ is given for three different numbers of particles:
$N = 64,128$ and $256$.  The corresponding value of the particle density, $na^3$,
for the $64$ particle case is shown on the top axis for reference.  At liquid
helium densities, the condensate fraction is roughly twice that of bulk liquid
$^4He$.  This difference can be understood by noting the shape of the radial
condensate density shown for the $a=256a_{Rb}$ case in Fig. 6.   Here, we
see that while the maximum particle density occurs at the center the trap,
the condensate density is peaked in the low density region at the edge of the
cloud.  This dilute region allows for a larger fraction of particles to
occupy the condensate orbital than in an uniform system at $^4He$ densities.

    \begin{figure}
\begin{center}
\includegraphics[width=3in]{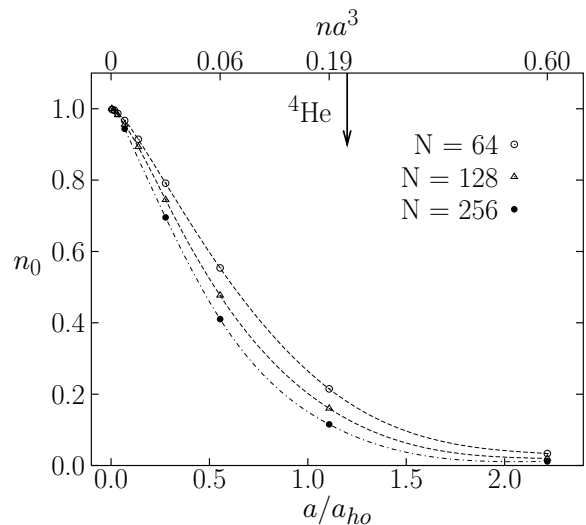}
 \caption{\footnotesize
Condensate fraction, $n_0$, at zero temperature for interacting
hard spheres in a harmonic trap as a function of the hard sphere
diameter, $a$, in units of the trap length, $a_{ho} =
\sqrt{{\hbar}/{m\omega_{_{ho}}}}$, for systems with $N = $ 64, 128 and
256 particles. The top axis indicates the corresponding values of
$n a^3$ found in the center of the trap for the 64 particle system. The arrow
indicates the value of $a/a_{ho}$ at which $n a^3$ is the
same as liquid $^4$He at SVP.
        }
\end{center}
    \end{figure}

Fig. 9 shows the effect of increased scattering length, $a$, on
the momentum distribution of particles in a isotropic harmonic trap at $T = 0$.
The values for the scattering length and the trap configuration
under consideration correspond to those shown for the spatial
distribution in Fig. 5.

    \begin{figure}
\begin{center}
\includegraphics[width=3in]{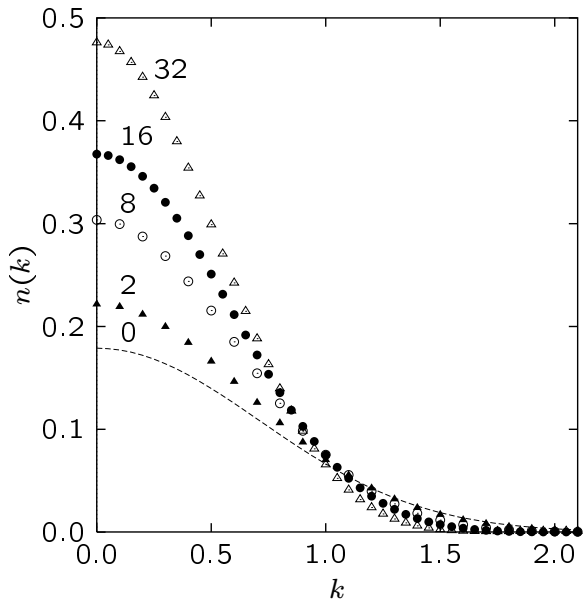}
 \caption{\footnotesize
Momentum distribution, $n(k)$, for 128 hard spheres with diameter $a = 0,1,..32$
times the scattering length of $^{87}Rb$, $a_{Rb} = 4.33 \times 10^{-3}$, in a
harmonic trapping potential.  Momentum is in units of the inverse of
the characteristic length of the trap $a^{-1}_{ho} = \sqrt{{m \omega_{_{ho}}}/{\hbar}}$.
        }
\end{center}
    \end{figure}
\section{Discussion}
In this section we compare the present MC values for the density, condensate
fraction and energy of bosons in a trap with mean field (MF) and Thomas-Fermi
(TF) approximation results.  The aim is to assess the limits of
applicability of the MF and TF expressions and to investigate the origin of any
differences between MF and MC values.  
We start with the density at the center of the
trap, $n(0)$, since this density is needed in MF expressions for the 
depletion of
the condensate and for the energy. 
Also, comparisons [4] of the density calculated in the TF
approximation and in mean field, Gross-Pitaevskii (GP), approximations show that
$n_{TF}(0)$ is accurate for large $Na/a_{ho}$ ($Na/a_{ho} \approx 100$ ).

The density of N independent bosons in an
asymmetric trap is 
$n_{ho}({\bf r}) = \phi_{ho}^2({\bf r}) = N/(\pi^{\frac{3}{2}}a_x a_y a_z) 
\exp[-(x^2/a_x^2+y^2/a_y^2+z^2/a_z^2)]$.  For the elliptical trap 
considered above,
$a_x=a_y=a_{ho}$ and $a_z = a_{ho}/ \sqrt{\lambda}$, the
density at the center of the trap is \cite{dalfovo99}
\begin{equation}
n_{ho}(0) = Na_{ho}^2a_z /\pi^{\frac{3}{2}} =
\sqrt{\lambda} N/ \pi^{\frac{3}{2}}a_{ho}^3
\label{d1}
\end{equation}
The $\phi_0(r)$ for a spherically
symmetric trap $(\lambda = 1)$ is shown in Fig.~1. As interaction is increased
(e.g. $a/a_{ho}$ is increased) the $\phi_0(r)$ spreads out and the density $n(0)$ at the
center of the trap decreases, as depicted in Fig.~5.  For $\lambda=1$, the
ratio of the density at the center with interaction to the $n_{ho}(0)$ for no
interaction in the TF approximation is
\begin{equation}
\frac{n_{TF}(0)}{n_{ho}(0)} = (\frac{15^{\frac{2}{5}}\sqrt{\pi}}{8})
(N a/a_{ho})^{-\frac{3}{5}} \label{density_ratio}.
\label{d2}
\end{equation}
This ratio calculated using MC, $n_{MC}(0)/n_{ho}(0)$, for increasing 
scattering length, $a/a_{ho} = F a_{Rb} = F \times 4.33 \times 10^{-3}$ with 
$F = 1,4,8,16,32 $ and $64$ can be obtained from  
Fig. 5 and is listed in Table 1.  The TF ratio agrees well with the MC
ratio in the range $4<F<16$ or $2 < N a/a_{ho} < 10$. 
We expect the TF result to be 
inaccurate at small $Na/a_{ho}$ because the TF limit is a large $Na/a_{ho}$ 
approximation. 
\begin{table}
\begin{center}
\begin{tabular}{c
                @{\hspace{.1cm}}c
                @{\hspace{.1cm}}c
                @{\hspace{.1cm}}c
                @{\hspace{.1cm}}c
                @{\hspace{.1cm}}
               }
\hline
\hline
\rule[-0.2cm]{0cm}{.5 cm}
$a/{a_{Rb}}$ & $Na/a_{ho}$ & $n_{MC}(0)a^3$ & $n_{TF}(0)/n_{ho}(0)$ &
${n_{MC}(0)}/{n_{ho}(0)}$ \\ 
\hline
\rule[-0.1cm]{0cm}{.5 cm}
1 & 0.55 & $1.3 \times 10^{-6}$ & 0.93 & 0.71  \\
4 & 2.22 & $4.3\times 10^{-5}$& 0.41 & 0.38\\
8 & 4.43 & $2.4\times 10^{-4}$ & 0.27 & 0.25\\
16 & 8.87 & $1.0\times 10^{-3}$ & 0.17 & 0.14\\
32 & 17.7 & $4.3\times 10^{-3}$ & 0.12 & 0.075\\
64 & 35.5 & $1.9\times 10^{-2}$ & 0.077 & 0.041\\
\hline
\hline
\end{tabular}
\caption{
The ratio of the density at the center of the trap calculated using the 
Thomas-Fermi expression, $n_{TF}(0)$, and in the present VMC evaluation, 
$n_{MC}(0)$, to the density for no interaction, $n_{ho}(0)$.  The ratio
decreases as the scattering length / hard core diameter $a/a_{Rb}$ is increased
as shown in Fig.~5.}
\end{center}
\end{table}

In addition, at large $Na/a_{ho}$, the density exceeds the dilute limit so
that all mean field theories become inaccurate.  The dilute limit is
exceeded at $n(0)a^3 \gapx 10^{-3}$ which corresponds to $F \geq 16$ for the
gas considered in Table~1.  
Between these limits however, for
$Na/{a_{ho}} \gapx 5$ and $n(0)a^3 \lapx 10^{-3}$, the $n_{TF}(0)$ gives a 
good estimate of the boson number density at the center of the trap.

For a uniform Bose gas, the MF, Bogoliubov theory predicts a condensate
fraction $n_0 = N_0/N = 1- (8/3)(na^3/\pi)^{\frac{1}{2}}$ where $n=N/V$
is the uniform density \cite{bogoliubov47}. That is, the fraction of bosons 
excited out of the condensate is
\begin{equation}
\frac{\delta N}{N} = \frac{N-N_0}{N} = 
\frac{8}{3}(\frac{na^3}{\pi})^{\frac{1}{2}}
\label{d3}.
\end{equation}
The corresponding result for bosons in a spherical harmonic trap is 
\cite{dalfovo99}
\begin{equation}
\frac{\delta N}{N} = \frac{5\pi}{8}(\frac{n_{TF}(0)a^3}{\pi})^{\frac{1}{2}}
\label{d4}.
\end{equation}
Fig.~7 shows that the mean field expressions (\ref{d3}) and (\ref{d4}) 
agree well with the 
present MC values of $n_0$ for $N=128$ bosons up to $a/a_{ho} \approx 16a_{Rb}$
or up to $n(0)a^3\approx 10^{-3}$.
Specifically, for $F=8$ at $n_{MC}(0)a^3 = 2.4 \times 10^{-4}$, the MC value for
the depletion in Fig.~7 is $\delta N/N = 1.9 \%$ while the Bogoliubov result 
from (\ref{d4}) is $\delta N/N = 1.7\%$.
For larger values of $n(0)a^3$ the depletion
is significantly underestimated by the Bogoliubov expressions.  This agrees
with the findings of MC determinations for $n_0$ in a uniform Bose gas\cite{giorgini}.
In the uniform case, (\ref{d3}) agrees with MC values up to 
$na^3 \approx 5\times 10^{-3}$ and
underestimates depletion at larger $na^3$.  The MF results are
accurate up to somewhat higher densities in the bulk probably because the density itself
does not depend on the interaction in a uniform Bose gas.  Recent experiments 
\cite{cornish00} have obtained stable condensates at 
densities corresponding to $na^3 \approx 10^{-2}$.  At this density, the condensate
fraction is $\approx .85$ and effects resulting from depletion are expected to be
significant. 

For a uniform Bose gas, the MF, Bogoliubov expression for the energy including
the leading correction arising from depletion is 
\cite{bogoliubov47}
\begin{equation}
\frac{E}{N} = 4\pi na^3[1+ \frac{128}{15}(\frac{na^3}{\pi})^{\frac{1}{2}}]
\label{d5}.
\end{equation}
The corresponding TF expression for bosons in a spherical trap ($\lambda$ = 1) is 
\cite{dalfovo99}
\begin{equation}
\frac{E}{N}=\frac{5}{7}\mu_{TF}[1+\frac{7\pi}{8}(\frac{n_{TF}(0)a^3}{\pi})
^{\frac{1}{2}}]
\label{d6}
\end{equation}
where $\mu_{TF} = \frac{1}{2}\hbar\omega_{\perp}
(15Na/a_{ho})^{\frac{2}{5}}$ 
and from (\ref{d1}) and (\ref{d2}) 
$n_{TF}(0)a^3=(15^{\frac{2}{5} }/8\pi)N^{\frac{2}{5} }(a/a_{ho})^{\frac{12}{5} }$

The above expressions may be used to estimate the energy of bosons in an anisotropic trap by
 replacing $a_{ho}$ with the geometric mean $a_g=(a_x a_y a_z)^{\frac{1}{3}}$
which is $a_g = a_{ho}\lambda^{-\frac{1}{6}}$ for the trap discussed in Fig.~2.
The energy of independent, non-interacting bosons in this anisotropic trap
is 
\begin{equation}
(E/N) \rightarrow E_{ho} = \hbar\omega_{ho}(1+\lambda/2) = \hbar\omega_{ho}(2.414)
\end{equation}
$(\omega_{ho} = \omega_{\perp})$.  
   \begin{figure}
\begin{center}
\includegraphics[width=3.25in]{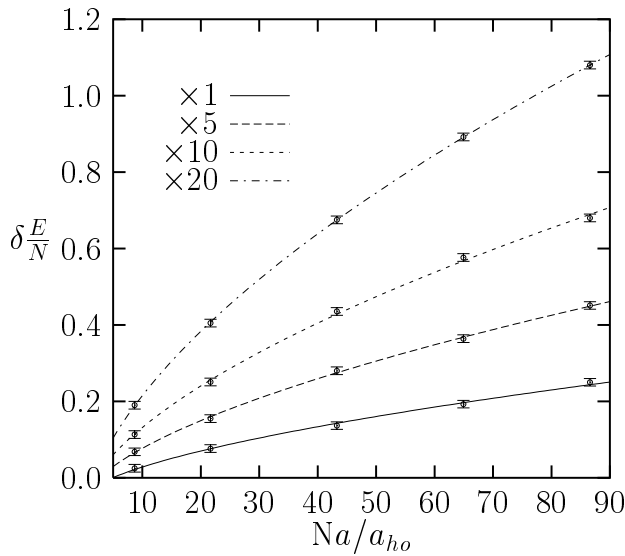}
\caption{Energy difference between present VMC and GP results,
$\delta{E/N}$, as a function of N$a/a_{ho}$ for four different
values of $a/a_{Rb} =$ 1, 5, 10, and 20.
The lines are fits of $\delta E/N = m(a/a_{ho})^{\frac{8}{5}}N^{\frac{3}{5}}+b$
to the data points.}
\end{center}
    \end{figure} 
In Fig.~2, ($E/N$) has the non-interacting boson limit at $N \rightarrow 0$.
As N increases, ($E/N$) increases.  As seen from (31), the mean field energy 
in the TF approximation $(E/N)_{TF} = 5\mu_{TF}/7$ increases 
with N as $(E/N)_{TF} \propto
(Na/a_{ho})^{\frac{2}{5} }$.  The $(E/N)_{MC}$ in Fig.~2 follows the dependence 
approximately.
However, direct evaluation of (31) shows that $(E/N)_{TF}$ underestimates the
energy significantly, by 25\% at N=20,000 (N$a/a_{ho}$ = 86.6).  Thus while the
Thomas Fermi density $n_{TF}(0)$ at the center of the trap is accurate, the
TF energy is a poor approximation in this density regime.  This is because the TF 
approximation underestimates the density at large $r$ and the high energy particles at 
large $r$ contribute significantly to the energy. 
\begin{table}[b!]
\begin{center}
\begin{tabular}{l@{\hspace{.65cm}}c@{\hspace{.65cm}}c@{\hspace{.65cm}}c@{\hspace{.65cm}}c}
\hline
\hline
\rule[-0.15cm]{0cm}{.5 cm}
$N$ & $Na/a_{ho}$ & $n_{TF}(0)a^3$ & $(\frac{\delta N}{N})_{MF}$
& $(\frac{\delta E}{E})_{MC}$ \\
\hline
\rule[-0.0cm]{0cm}{.3 cm}
5000 & 21.6 & $1.1\times 10^{-5}$ & 0.37\% & 1.1\%\\
10000 & 43.3 & $1.5\times 10^{-5}$ & 0.43\% & 1.8\%\\
20000 & 86.6 & $2.0\times 10^{-5}$ & 0.50\% & 2.5\%\\
\hline
\hline
\end{tabular}
\caption{Depletion of the condensate $(\delta N/N)_{MF}$ calculated using
Mean Field expansions (\ref{d4}) for the Bose gas considered in Fig.~2 for
N$=5000$,$10,000$, and $20,000$.  The $(\delta E/E)_{MC}$ is the fractional difference
between the present MC energy of the gas and the GP \cite{dalfovo99} energy
of the condensate, $\delta E = E_{MC} - E_{GP}$.  This
difference is consistent with the depletion of the condensate and mean field 
expressions of $\delta E/E$ arising from depletion.}
\end{center}
\end{table}

In Fig.~2 we see that the present
Monte Carlo $E/N$ lies above the Gross-Pitaevskii $E/N$, by $2.5\%$ at
$N = 20,000$.  A difference could arise for two reasons.  

Firstly, the MC energy is
the E/N of the whole Bose gas while the GP equation calculates the 
$E/N$ of the condensate only. 
To investigate the effects of depletion on $E/N$ we note, comparing (28) and
(30), that the fractional change in energy arising from depletion, $\delta E/E$,
can be related to the fraction of atoms excited out of the condensate as,
\begin{equation}
\delta E/E = (128/15)(na^3/\pi)^{\frac{1}{2} } = (16/5)\delta N/N
\end{equation}
in the bulk.
The corresponding equation for the trap is, from (29) and (31), 
\begin{equation}
\delta E/E = (7/5)\delta N/N.  
\end{equation}
These are lowest order expressions.
In Table~2 we list the depletion of the condensate 
$\delta N/N$ for the 
bosons in the anisotropic trap considered in Fig.~2 for $N = 5000, 10,000$
and $20,000$ predicted by (\ref{d4}).  
We expect these predictions to be accurate 
since $n_{TF}(0)$ is reliable and $n_{TF}(0)a^3$ is small.  
The interaction 
depletes the condensate $0.5\%$ at $N=20,000$.
From the above connection between $\delta E$ and $\delta N$, the energy $(E/N)$
including depletion is expected to lie 0.7-1.6\% above the energy of the condensate.
On this basis, VMC and GP energies are consistent.

To explain the connection between the difference in VMC and GP energies and 
depletion of the condensate more fully, we have plotted 
$\delta (E/N) = (E_{MC}-E_{GP})/N$ vs $Na/a_{ho}$ in Fig.~10.  From (31), the
difference from the mean field energy arising from depletion
is 
\begin{equation}
\begin{array}{rl}
\delta (E/N)_{TF} = & (5\pi\mu_{TF}/8)(n_{TF}(0)a^3/\pi)^{\frac{1}{2} }\\ 
\propto & N^{\frac{3}{5} } (a/a_{ho})^{\frac{8}{5} }
\end{array}
\end{equation}
in the TF limit.
In Fig.~10, we show the difference in the VMC and GP energies, $\delta (E/N)$, verses
the product $Na/a_{ho}$.  This difference is obtained from Fig.~4 for four values of 
$a/a_{ho} = 1,5,10,20$.  The lines in Fig.~10 are fits of
$\delta(E/N) = m N^{\frac{3}{5} }(a/a_{ho})^{\frac{8}{5} } + b$ to the data where
the fitting parameters $m$ and $b$ where allowed to change for each $a/a_{ho}$ value.
The good fit of these lines shows that $\delta(E/N)$ reflects the dependence 
on N expected for a difference in energy arising from exciting bosons out of
the condensate.  Thus the difference in $(E/N)_{MC}$ from $(E/N)_{GP}$ is consistent
in magnitude and dependence on N and $a/a_{ho}$ with that expected for an
$(E/N)_{MC}$ that includes bosons both in and above the condensate while $(E/N)_{GP}$
is the energy of bosons in the condensate only.
Thus we believe the MC and GP energies differ chiefly
because the MC includes ``excited" particles while the GP energy does not. 

Secondly, the present VMC $E/N$ is a genuine upper bound for the whole gas 
and could lie above the whole gas energy.  We have not 
tested the sensitivity of $E/N$ to different choices of the trial wave function.
In addition, while the present pair Jastrow function (\ref{jastrow_func}) is exact in the dilute
limit it does not contain any variational parameters and is therefor not optimized
for trapped hard spheres at higher densities.  As a result, at least some of 
the difference in $(E/N)_{MC}$ from $(E/N)_{GP}$ could arise from the present
choice of trial wave function.
For example, Fabrocini and Polls (FP) have evaluated the whole $E/N$ for 
bosons in a spherical trap
$a_x=a_y=a_z=4.33\times10^{-3}a_{ho}$ using correlated basis function
and hypernetted-chain (HNC) methods\cite{fabrocini99}.  
Both methods provide estimates of $E/N$ which lie below the present VMC $E/N$.
At $N=10^5$ the density ($n(0)a^3=2.5\times10^{-5}$) in this trap is similar
to that for $N=2\times10^4$ in the present elliptical trap.  At this density
the HNC $E/N$ lies 0.8\% above the GP energy of the condensate while the present
VMC $E/N$ is 2.5\% above GP.
FP also consider a mean field model incorporating quantum depletion which predicts 
an increase in $E/N$ of 1.2\% above the GP result at this density.
Generally, the HNC energy in the trap lies below the energy
expected for depletion and may be too low.  For example, the HNC energy for a uniform
gas of bosons lies above the energy expected for depletion.  Definite resolution of
these differences awaits a model independent evaluation of $E/N$ by Diffusion
Monte Carlo methods.

Finally, an important result of the present MC evaluation is that as the boson density $na^3$
increases, the condensate gradually moves from the center of the trap to the edges of the
trap as shown in Fig. 6.  At large $na^3$, the condensate is at the edges of the trap.
In this limit, the depletion of the condensate is large and the condensate seeks the regions
of lowest total density which are at the surface of the trap. Both the condensed
and uncondensed atoms must be included in the calculation to obtain this effect.  
This result is consistent with the calculations in liquid $^4$He droplets \cite{lewart88} 
which find the condensate concentrated at the surface of the droplet.
\acknowledgments
{
Stimulating discussions with Charles W. Clark and Timothy Ziman 
are gratefully acknowledged. 
J.L.D. acknowledges support from DE-SGGFP NASA Grant No. NGT5-40024. 
}


\begin{thebibliography}{999}%\parskip=0pt\itemsep 0pt%
\footnotesize

\bibitem {anderson95}%1
M.H. Anderson, J.R. Ensher, M.R. Matthews, C.E. Wieman, and E.A. Cornell, {\em Science} {\bf
269}, 198 (1995).

\bibitem{davis95}%2
K.B. Davis, M.-O. Mewes, M.R. Andrews, N.J. van Druten, D.S. Durfee, D.M. Kurn, and W. Ketterle,
{\em Phys. Rev. Lett.} {\bf 75}, 3969 (1995).

\bibitem{bradley95}%3
C.C. Bradley, C.A. Sackett, J.J. Tolett, and R.G. Hulet, {\em Phys. Rev. Lett.} {\bf 75}, 1687
(1995); C.C. Bradley, C.A. Sackett, and R.G. Hulet, {\em ibid.} {\bf 78}, 985 (1997).

\bibitem{dalfovo99}%4
F. Dalfovo, S. Giorgini, L. Pitaevskii, and S. Stringari, {\em Rev. Mod. Phys.} {\bf 71}, 463
(1999).

\bibitem{bogoliubov47}%5
N.N. Bogoliubov, {\em J. Phys. (Moscow)} {\bf 11}, 23 (1947).

\bibitem{gross61}%6
E.P. Gross, {\em Nuovo Cimento} {\bf 20}, 454 (1961).

\bibitem{pitaevskii61}%7
L.P. Pitaevskii, {\em Ah. Eksp. Teor. Fiz.} {\bf 40}, 646 (1961) [{\em Sov. Phys. JETP} {\bf 13},
451 (1961)].

\bibitem{hutchinson97}%8
D.A.W. Hutchinson, E. Zaremba, and A. Griffin, {\em Phys. Rev. Lett.} {\bf 78}, 1842 (1997).

\bibitem{krauth96}%9
W. Krauth, {\em Phys. Rev. Lett.} {\bf 77}, 3695 (1996).

\bibitem{gruter97}%10
P. Gr\"{u}ter, D. Ceperley, and F. Lalo\"{e}, {\em Phys. Rev. Lett.} {\bf 79}, 3549 (1997).

\bibitem{holzmann99}%11
M. Holzmann, W. Krauth, and M. Naraschewski, {\em Phys. Rev. A} {\bf 59}, 2956 (1999).

\bibitem{giorgini}%12
S. Giorgini, J. Boronat and J. Casulleras, {\em Phys. Rev. A} {\bf 60}, 5129 (1999). 

\bibitem{lewart88}%13
D.S. Lewart, V.R. Pandharipande, and S.C. Pieper, {\em Phys. Rev. B} {\bf 37}, 4950 (1988).

\bibitem{gardner95}%14
J.R. Gardner et al. {\em Phys. Rev. Lett.} {\bf 74}, 3764 (1995).

\bibitem{kalos74}%15
M.H. Kalos, D. Levesque, and L. Verlet, {\em Phys. Rev. A} {\bf 9}, 2178 (1974).

\bibitem{baym76}%16
G. Baym, {\em Lectures on Quantum Mechanics},425,(W. A. Benjamin, Inc.,1976).

\bibitem{wilks67}%17
J. Wilks, {\em The Properties of Liquid and Solid Helium} (Clarendon Press, Oxford, 1967).

\bibitem{dalfovo96}%18
F. Dalfovo and S. Stringari, {\em Phys. Rev. A} {\bf 53}, 2477 (1996).

\bibitem{javanainen96}%19
J. Javanainen, and S.M. Yoo, {\em Phys. Rev. Lett.} {\bf 76}, 161 (1996).

\bibitem{lowdin55}%20
P.O. L\"owdin, Phys. Rev. {\bf 97}, 1474 (1955).

\bibitem{onsager56}%21
L. Onsager and O. Penrose, Phys. Rev. {\bf 104}, 576 (1956).

\bibitem{huang63}%22
K. Huang, {\em Statistical Mechanics},P. 276,(Wiley \& Sons,1963).

\bibitem{metropolis}%23
N. Metropolis, A.E. Rosenbluth, M.N. Rosenbluth, 
A.H. Teller and E. Teller , J. Chem. Phys. {\bf 21}, 1087 (1953).

\bibitem{ziegler97}
K. Ziegler and A. Shukla, {\em Phys. Rev. A} {\bf 56}, 1438 (1997). 

\bibitem{vmc_Review}
M.P. Nightingale and C.J. Umrigar, 
{\em Quantum Monte Carlo Methods in Physics and Chemistry},P. 129,(Kluwer,1999).

\bibitem{jastrow}
R. Jastrow. {\em Phys. Rev.} {\bf 98} 1479 (1955).

\bibitem{pollock92}
%FINITE-SIZE-SCALING ANALYSIS OF A SIMULATION OF THE HE-4 SUPERFLUID TRANSITION
E.L. Pollock and  K.J. Runge,{\em Phys. Rev. B} {\bf 46} 3535 (1992)

\bibitem{cornish00}
S.L. Cornish, N.R. Claussen, J.L. Roberts, E.A. Cornell, C.E. Wieman, {\em http://xxx.lanl.gov/abs/cond-mat/0004290}


\bibitem{fabrocini99}
A. Fabrocini and A. Polls, {\em Phys. Rev. A} {\bf 60}, 2319 (1999).

\end{thebibliography}
\end{document}